\documentclass[onecolumn,preprintnumbers,eqsecnum,superscriptaddress,10pt,floatfix,amsmath,amssymb]{revtex4}
\usepackage{float}
\usepackage{graphicx}
\usepackage{epsf}
\usepackage{dcolumn}   
\usepackage{bm}        
\usepackage{color}
\usepackage{appendix}

\newcommand{\be}{\begin{equation}}
\newcommand{\en}{\end{equation}}
 \newcommand{\bea}{\begin{eqnarray}}
 \newcommand{\ena}{\end{eqnarray}}

\begin{document}

\title{Divergence Behavior of Thermodynamic Curvature Scalar at Critical Point in the Extended Phase Space of Generic Black Holes}

\author{Ya-Peng Hu}\email{huyp@nuaa.edu.cn}
\address{College of Science, Nanjing University of Aeronautics and Astronautics, Nanjing 210016, China}
\address{Key Laboratory of Aerospace Information Materials and Physics (NUAA), MIIT, Nanjing 211106, China}
\address{Center for Gravitation and Cosmology, College of Physical Science and Technology, Yangzhou University, Yangzhou 225009, China}

\author{Liang Cai}\email{cailiang0531@nuaa.edu.cn}
\address{College of Science, Nanjing University of Aeronautics and Astronautics, Nanjing 210016, China}

\author{Xiao Liang}\email{xliang@nuaa.edu.cn}
\address{College of Science, Nanjing University of Aeronautics and Astronautics, Nanjing 210016, China}

\author{Shi-Bei Kong}\email{kongshibei@nuaa.edu.cn}
\address{College of Science, Nanjing University of Aeronautics and Astronautics, Nanjing 210016, China}

\author{Hongsheng Zhang}\email{sps_zhanghs@ujn.edu.cn}
\address{School of Physics and Technology, University of Jinan, 336 West Road of Nan Xinzhuang, Jinan, Shandong 250022, China}

\begin{abstract}
The $P$-$V$ phase transition and critical behavior in the extended phase space of asymptotic Anti-de Sitter (AdS) black holes have been widely investigated, in which four critical exponents around critical point are found to be consistent with values in the mean field theory. Recently, another critical exponent $\nu$ related to divergent correlation length at critical point is proposed by using thermodynamic curvature scalar $R_N$ in the charged AdS black hole. In this paper, we develop a method to investigate the divergent behavior of $R_N$ at critical point, and find that the divergent behavior of $R_N$ around the critical point expresses a universal property in generic black holes. We further directly apply this method to investigate black holes in de Rham-Gabadadze-Tolley (dRGT) massive gravity to check this universality. Those results shed new lights on the microscopic properties of black holes.


\end{abstract}

\maketitle



\section{Introduction}

In the study of the black hole thermodynamics, numerous examples of phase transitions have been found since the discovery of the famous Hawking-Page phase transition \cite{HP} of the Schwarzschild-AdS (Anti-de Sitter) black holes, such as \cite{Cvetic:1999ne,Cvetic:1999rb}. Among these, $P$-$V$ phase transitions
\cite{Caldarelli:1999xj,Kastor:2009wy,Dolan:2010ha,Dolan:2011xt,Dolan:2011jm,Cvetic:2010jb,Lu:2012xu} have also been found
in the extended phase spaces of the (asymptotic) AdS black holes in recent years, which are proved to be a powerful tool in the exploration of the AdS black holes \cite{Gunasekaran:2012dq,Hendi:2012um,Cai:2013qga,Majhi:2016txt,Miao:2018fke,Spallucci:2013osa,Wei:2012ui,Xu:2015rfa,Zhao:2013oza,Hu:2018qsy,Banerjee:2010bx}.
The discovery of the $P$-$V$ phase transitions in AdS black holes are based on the introduction of the thermodynamic pressure in the black hole system inspired by the studies of the acceleration in cosmology, and the thermodynamic pressure in the AdS spacetime is provided by the negative cosmological constant $\Lambda$ via
\begin{eqnarray}
	P=-\dfrac{\Lambda}{8\pi}=\dfrac{\left(n-1\right)\left(n-2\right)}{16\pi {l}^2},
\end{eqnarray}
where $l$ is the AdS radius and $n$ is the spacetime dimension. Combined with the first law of black hole thermodynamics, one can show that the $P$-$V$ phase transitions in the extended phase space of AdS black holes are very similar to the gas-liquid phase transitions in the van der Waals (vdW) system. In the subsequent studies, the $P$-$V$ phase transitions are also discovered in various modified theories of gravity \cite{Hendi:2012um,Spallucci:2013osa,Cai:2013qga,Xu:2015rfa,Majhi:2016txt,Hu:2018qsy} beside Einstein gravity. Furthermore, one can also extract the four critical exponents near the critical points \cite{Kubiznak:2012wp} of the black hole $P$-$V$ phase transitions. This provides further insights into the microscopic properties of the black holes.

Thermodynamic geometry is another useful and convenient way to study the thermodynamic properties of the black holes\cite{Ruppeiner:1995zz,Weinhold,Wei:2019uqg,Wei:2019ctz,Wei:2020poh}, which has a close relationship with phase transitions. Weinhold first introduced the thermodynamic geometry in the equilibrium thermodynamic system \cite{Weinhold}. Ruppeiner improved his theory by introducing another thermodynamic geometry (now called Ruppeiner geometry) from the Boltzmann's entropy formula~\cite{Ruppeiner:1995zz}, where the distance between two neighbouring fluctuation states is described by a line element $dl^2$. In the Ruppeiner geometry, there is a Ruppeiner curvature scalar $R$
analog to the Ricci scalar in Riemann geometry. It is related to the correlation length $\xi$ of the thermodynamic system via \cite{Ruppeiner:1995zz}
\begin{eqnarray}
	R\sim \kappa \xi^{\bar{d}},
\end{eqnarray}
where $\kappa$ is a constant and $\bar{d}$ is defined as the physical dimensionality of the system.
Note that, the Ruppeiner curvature scalar provides information about the phase transitions of the black holes, and the correlation length $\xi$ becomes divergent when the thermodynamic system approaches to the critical temperature of phase transitions. Near the critical point, the correlation length $\xi$ is related to the distance parameter $t$ (relative to the critical point) by $\xi \sim t^{-\nu}$, which provides a new critical exponent $\nu$. It is very interesting that a recent article \cite{Wei:2019uqg} presents this new critical exponent $\nu$ in the $P$-$V$ criticality of the charged AdS black hole by using Ruppeiner thermodynamic geometry. This new critical exponent provides new clues to understand the microstructure of black holes. Furthermore, this work also shows a novel universal property of charged AdS black hole in Einstein gravity firstly, where the divergence behavior of Ruppeiner curvature scalar is characterized by a dimensionless constant $-1/8$ that is identical to the van der Waals fluid~\cite{Wei:2019uqg}. Later, this universal property with dimensionless constant $-1/8$ is also confirmed in Einstein-Gauss-Bonnet gravity~\cite{Wei:2019ctz,Wei:2020poh}. Therefore, it is necessary to further investigate the divergence behavior of Ruppeiner curvature scalar in more cases, and further verify this universality with dimensionless constant $-1/8$. In this paper, we develop a non-trivial method to study the behavior of Ruppeiner curvature scalar around the critical point, and demonstrate this universality in generic black holes.

The de Rham-Gabadadze-Tolley (dRGT) massive gravity is a well-defined and leading modified gravity with massive gravitons, which has been widely investigated in various issues. Moreover, the $P$-$V$ criticality in the extended phase space of asymptotic AdS black holes in dRGT massive gravity has presented special properties different from other theories of gravity \cite{Xu:2015rfa}. For example, $P$-$V$ criticality is not only found in the black hole system with spherical horizon ($k=1$), but also found in the Ricci flat or planar horizon case ($k=0$) and hyperbolic horizon case ($k=-1$), where $k$ characterizes the horizon curvature. Therefore, a natural and interesting question is to study the divergence behavior of Ruppeiner curvature scalar in dRGT massive gravity, and check whether the dimensionless constant $-1/8$ is still universal or not. Through using our method, we check that the dimensionless constant $-1/8$ is also universal in dRGT massive gravity case, where the dimensionless constant is indeed independent of the graviton mass and horizon topology.

The rest of the paper is organized as follows. In Sec.II, we briefly review the Ruppeiner thermodynamic geometry and the thermodynamic curvature scalar $R_N$. In Sec.III, we propose a non-trivial method to investigate the divergence behavior of Ruppeiner curvature scalar around the critical point, and find that the dimensionless constant $-1/8$ indeed expresses a universal property in generic black holes. In Sec.IV, we make a solid demonstration of this universality with more general expansions. In Sec.V, we use our method to check this universality in the four-dimensional dRGT massive gravity case. In the last section, we summarize our work and make some discussions.

\section{Warm-up: Ruppeiner thermodynamic geometry and curvature scalar $R_N$}\label{ru}
In this section, we make a brief review on some main results of Ruppeiner thermodynamic geometry and definition of thermodynamic curvature scalar $R_N$ as a warm-up. In Ruppeiner thermodynamic geometry, the distance between two neighbouring fluctuation states is described by a line element $dl^2$, which reads\cite{Ruppeiner:1995zz,Wei:2019uqg}
\begin{eqnarray}
dl^2=-\dfrac{\partial ^2S}{\partial x^\mu\partial x^\nu}dx^\mu dx^\nu,	
\end{eqnarray}
where $S$ is the entropy of a small subsystem, and $x^\mu$ are independent thermodynamic variables of the whole system, i.e. supposed to be conservative and additive fluctuating parameters. This line element is just the well-known Ruppeiner thermodynamic information geometry.

In Ruppeiner thermodynamic geometry corresponding to the above line element $dl^2$, the curvature scalar $R$ can be obtained. If one chooses the thermodynamic internal energy $U$ and volume $V$ as fluctuation variables, the line element is expressed as
\begin{eqnarray}
dl^2=-\dfrac{\partial ^2S}{\partial U^2}dU^2-2\dfrac{\partial^2S}{\partial U\partial V}dUdV-\dfrac{\partial^2S}{\partial V^2}dV^2.
\end{eqnarray}
For a canonical ensemble, temperature $T$ and thermodynamic volume $V$ are offen used as two independent variables of a system. Therefore, the line element $dl^2$ needs to be re-expressed by new variables $(T,V)$. From the first law of the thermodynamic system $dS=\dfrac{1}{T}dU+\dfrac{P}{T}dV$, one leads to
\begin{eqnarray}
\left(\dfrac{\partial S}{\partial U}\right)_V=\dfrac{1}{T},~~\left(\dfrac{\partial S}{\partial V}\right)_U=\dfrac{P}{T}.
\end{eqnarray}
Hence,
\begin{eqnarray}
d\left(\dfrac{1}{T}\right)&=&\left(\dfrac{\partial^2 S}{\partial U^2}\right)_V dU+\left(\dfrac{\partial^2 S}{\partial U \partial V}\right)dV,~~d\left(\dfrac{P}{T}\right)=\left(\dfrac{\partial^2S}{\partial V^2}\right)_UdV+\left(\dfrac{\partial^2S}{\partial U\partial V}\right)dU,
\end{eqnarray}
and the line element $dl^2$ is further equal to
\begin{eqnarray}
dl^2&=&-d\left(\dfrac{1}{T}\right)dU-d\left(\dfrac{P}{T}\right)dV=\dfrac{1}{T^2}dTdU-\frac{1}{T}dPdV+\frac{P}{T^2}dTdV.\\\nonumber
\end{eqnarray}
Using $dU=C_VdT+\left[T(\frac{\partial P}{\partial T})_V-P\right]dV$ and $dP=(\frac{\partial P}{\partial T})_VdT+(\frac{\partial P}{\partial V})_TdV$ with some simple calculations, one can obtain the line element $dl^2$ with $(T,V)$ variables as \cite{Wei:2019uqg}
\begin{eqnarray}
dl^2&=&\dfrac{C_V}{T^2}dT^2-\dfrac{\left(\partial_V P\right)_T}{T}dV^2. \label{Newelement}
\end{eqnarray}
As a consequence, the curvature scalar $R$ is expressed as
\begin{eqnarray}
R&=&\dfrac{1}{2C^2_V \left(\partial _VP\right)^2}[T(\partial _VP)(\partial _VC_V)^2+T(\partial _VP)(\partial _TC_V)(-\partial _VP+T\partial _{T,V}P)+C_V(\partial _VP)^2\nonumber \\
& &+TC_V(\partial _VC_V)(\partial _{V,V}P)-T^2C_V(\partial _{T,V}P)^2-2T C_V(\partial _VP)(\partial _{V,V}C_V)+2T^2C_V(\partial _VP)(\partial _{T,T,V}P)].\label{R}
\end{eqnarray}
When investigating $P$-$V$ criticality in extended phase space of asymptotic AdS black holes, the heat capacity $C_V$ is often zero because the entropy $S$ of a black hole is usually the function of horizon radius $r_h$, i.e. $S(r_h)$, while thermodynamic volume is $V=\frac{4 \pi}{3}r_h^3$. Therefore, the usual definition of curvature scalar $R$ in (\ref{R}) will be divergent. In order to avoid this divergence, a new normalized curvature scalar $R_N$ is defined as $R_N=R C_V$ with heat capacity $C_V$ approaching to zero. This normalized thermodynamic curvature scalar $R_N$ with zero heat capacity case is further simplified as \cite{Wei:2019uqg}
\begin{eqnarray}
R_N=\frac{\left(\frac{\partial P}{\partial V}\right)^2_T-T^2\left(\frac{\partial^2 P}{\partial T\partial V}\right)^2+2T^2\left(\frac{\partial P}{\partial V}\right)_T\left(\frac{\partial^3 P}{\partial T^2\partial V}\right)}{2\left(\frac{\partial P}{\partial V}\right)^2_T},\label{RN}
\end{eqnarray}
and $R_N$ is a more penetrating probe to investigate the microscopic properties of black holes.

\section{ Divergence Behavior of the Thermodynamic Curvature Scalar $R_N$ at the Critical Point in generic black holes }\label{four}
For the above definition of thermodynamic curvature scalar $R_N$, it has exhibited a divergence behavior at the critical point of $P$-$V$ criticality in the extended phase space of the charged AdS black hole, while this divergence behavior is characterized by a dimensionless constant $-1/8$. Moreover, this dimensionless constant is identical to that for a van der Waals fluid, which implicates a universal behavior \cite{Wei:2019uqg}. In this section, we propose a non-trivial method to further study the divergence behavior of thermodynamic curvature scalar at the critical point in generic black holes, and check this universality.

Since we do not know the explicit form of equation states for generic black holes, we just start from the general equation of state $P=P(T,V)$ of a black hole. The thermodynamic pressure $P$ of all known AdS black holes obeys such a formula. On the other hand, if black holes exist $P$-$V$ criticality, i.e. the critical temperature $T_c$ and volume $V_c$ can be calculated from $\left[\left(\dfrac{\partial P}{\partial V}\right)_T\right]_c=\left[\left(\dfrac{\partial ^2P}{\partial V^2}\right)_T\right]_c=0$, we can expand $P$ around the critical point
\begin{eqnarray}
P&=&P_c+\left[\left(\dfrac{\partial P}{\partial T}\right)_V\right]_c\left(T-T_c\right)+\left[\left(\dfrac{\partial^2 P}{\partial V\partial T}\right)\right]_c\left(T-T_c\right)\left(V-V_c\right)+\dfrac{1}{3!}\left[\left(\dfrac{\partial^3 P}{\partial V^3}\right)_T\right]_c\left(V-V_c\right)^3+\nonumber
\\& &\dfrac{1}{2!}\left[\left(\dfrac{\partial^2 P}{\partial T^2}\right)_V\right]_c\left(T-T_c\right)^2+\frac{3}{3!}\left[\frac{\partial^3 P}{\partial T\partial^2 V}\right]_c\left(T-T_c\right)\left(V-V_c\right)^2+\dfrac{1}{4!}\left[\left(\dfrac{\partial^4 P}{\partial V^4}\right)_T\right]_c\left(V-V_c\right)^4... ,\label{Pressure}
\end{eqnarray}
where the $P_c$ is the corresponding critical pressure. In order to simplify calculations, we obtain the reduced pressure $\tilde{P}=P/P_c$ as
\begin{eqnarray}
\tilde{P}&=&1 + R\left(\tilde{T} - 1\right) + B\left(\tilde{T} - 1\right)\left(\tilde{V} - 1\right) + D\left(\tilde{V} - 1\right)^3 + H\left(\tilde{T} - 1\right)^2\nonumber
\\& & +I\left(\tilde{T} - 1\right)\left(\tilde{V} - 1\right)^2 + G\left(\tilde{V} - 1\right)^4 + ..., \label{PressureGeneral}
\end{eqnarray}
where $\tilde{T}=T/T_c,\tilde{V}=V/V_c$ and
\begin{eqnarray}
R&=&\dfrac{T_c}{P_c}\left[\left(\dfrac{\partial P}{\partial T}\right)_V\right]_c,~ B=\dfrac{T_c V_c}{P_c}\left[\left(\dfrac{\partial^2 P}{\partial T \partial V}\right)_V\right]_c,~ D=\dfrac{V^3_c}{6P_c}\left[\left(\dfrac{\partial^3 P}{\partial V^3}\right)_T\right]_c,~H=\dfrac{T^2_c}{2P_c}\left[\left(\dfrac{\partial^2 P}{\partial T^2}\right)_V\right]_c,\nonumber
\\I&=&\frac{T_c V^2_c}{2P_c}\left[\frac{\partial^3 P}{\partial T\partial^2 V}\right]_c,~G=\dfrac{V^4_c}{24P_c}\left[\left(\dfrac{\partial^4 P}{\partial V^4}\right)_T\right]_c.
\end{eqnarray}
Note that, the heat capacity $C_V$ of most black holes is zero or constant, which is just the case discussed in (\ref{RN}) for the normalized thermodynamic curvature scalar $R_N$. In addition, one will obtain the same result, if one uses the reduced quantities like the reduced pressure in (\ref{PressureGeneral}) to calculate $R_N$ in (\ref{RN}). Therefore, after substituting (\ref{PressureGeneral}) into the normalized thermodynamic curvature scalar $R_N$ in (\ref{RN}), we obtain
\begin{eqnarray}
R_N=\frac{\left[B (\tilde{T}-1)+3 D (\tilde{V}-1)^2+2 I (\tilde{T}-1) (\tilde{V}-1)+4 G (\tilde{V}-1)^3\right]^2-\tilde{T}^2 \left[B+2 I (\tilde{V}-1)\right]^2}{2 \left[B (\tilde{T}-1)+3 D (\tilde{V}-1)^2+2 I (\tilde{T}-1) (\tilde{V}-1)+4 G (\tilde{V}-1)^3\right]^2}. \label{RNGeneral}
\end{eqnarray}
Clearly, $R_N$ diverges at the critical point because both $\tilde{T}$ and $\tilde{V}$ approach to $1$.
	
	In order to investigate this divergence behavior of $R_N$ in details, we need to further find the relationship between $\tilde{V}$ and $\tilde{T}$ around the critical point. To make calculations more convenient, we rewrite $\tilde{P}$ as
	\begin{eqnarray}
	\tilde{P}=1-Rt-Btw+Dw^3+Ht^2-Itw^2+Gw^4+..., \label{Pyuehua}
	\end{eqnarray}
where $t=1-\tilde{T}$, $w=\tilde{V}-1$. From (\ref{Pyuehua}), we obtain $d\tilde{P}=\left(-Bt+3Dw^2-2Itw+4Gw^3+...\right)dw$, where $t$ is constant in the isothermal curve. Note that, below the critical temperature, there are two phases for generic black hole system, i.e. small and large black hole phases.
Therefore, we denote the small solution as $w_s$ and $\tilde{V}_s$, the large solution as $w_l$ and $\tilde{V}_l$. By using the Maxwell's equal area law $\tilde{P}^*\left(\tilde{V}_s-\tilde{V}_l\right)=\int^s_l\tilde{P}d\tilde{V}$ in the $P$-$V$ phase diagram\cite{Majhi:2016txt}, we further obtain
\begin{eqnarray}
	\int^{w_s}_{w_l}-w\left(-Bt+3Dw^2-2Itw+4Gw^3+...\right)dw+\int^{w_s}_{w_l}\left(-Bt+3Dw^2-2Itw+4Gw^3+...\right)dw=0.
	\label{MaxwellEqgen}
\end{eqnarray}
where $\tilde{P}^*$ is just the corresponding reduced pressure in the straight line with two coexistence phases. Note that, the second integral of (\ref{MaxwellEqgen}) should be zero just like the end point of vapor and the starting point of liquid have the same pressure, i.e. $\tilde{P}^*=\tilde{P}_l=\tilde{P}_s$, which implies
	\begin{eqnarray}
	B\left(w_l-w_s\right)t-D\left(w^3_l-w^3_s\right)+I\left(w^2_l-w^2_s\right)t-G\left(w^4_l-w^4_s\right)...=0, \label{V-T Eq1}
	\end{eqnarray}
	and thus from (\ref{MaxwellEqgen}), another result is
	\begin{eqnarray}
	 \frac{1}{2}B\left(w^2_l-w^2_s\right)t-\frac{3}{4}D\left(w^4_l-w^4_s\right)+\frac{2}{3}I\left(w^3_l-w^3_s\right)t-\frac{4}{5}G\left(w^5_l-w^5_s\right)...=0. \label{V-T Eq2}
	\end{eqnarray}
	Solving the above two equations (\ref{V-T Eq1}) and (\ref{V-T Eq2}), we can exactly obtain analytical reduced volumes of these two black hole phases, i.e. small and large black holes around the critical point, which are
\begin{eqnarray}
\tilde{V}_s=1-\sqrt{\frac{B}{D}} t^\frac{1}{2}+\left(\frac{I}{3D}-\frac{6BG}{15D^2}\right)t...,~~\tilde{V}_l=1+\sqrt{\frac{B}{D}} t^\frac{1}{2}+\left(\frac{I}{3D}-\frac{6BG}{15D^2}\right)t.... \label{FinalV}
\end{eqnarray}
where $B$ and $D$ are not zero. Therefore, after inserting (\ref{FinalV}) into (\ref{RNGeneral}), we obtain divergence behavior of thermodynamic curvature scalar $R_N$ around the critical point in generic black holes
\begin{eqnarray}
R_N\left(SBH\right)&=&-\frac{1}{8 t^2}+\frac{(5 D I-2 B G) \sqrt{\frac{B}{D}} }{10 B D t^{3/2}}+...,~~ R_N\left(LBH\right)=-\frac{1}{8 t^2}-\frac{(5 D I-2 B G) \sqrt{\frac{B}{D}} }{10 B D t^{3/2}}+...,\label{RNfinall}
\end{eqnarray}
for small(SBH) and large(LBH) black hole cases, respectively. The above result explicitly shows a universal divergent property, i.e. its divergence behavior is characterized by a dimensionless constant $-1/8$ in generic black hole system with non-zero $B$ and $D$.

\section{A solid demonstration of the universality with more general expansions}
Although we obtain the above results in (\ref{RNfinall}), it should be pointed out that those results seem to be underlying dependent on the higher-order terms of analytical reduced volumes in (\ref{FinalV}). Therefore, it is necessary to investigate whether the higher order terms of analytical reduced volumes in (\ref{FinalV}) affect the results of $R_N$ in (\ref{RNfinall}). In this section, we will make more general expansions to obtain the higher order terms of $\tilde{V_s}$ and $\tilde{V_l}$ with $t^{3/2}$ term, and give a solid demonstration of the universality.

We start from more general expansions of pressure around critical point, and take the expanding form of reduced pressure $\tilde{P}$ as
\begin{eqnarray}
	\tilde{P}&=&1+\sum_{i\geq 1}a_{i0}t^{i}+\sum_{j\geq 3}a_{0j}w^{j}+\sum_{i,j\geq 1}a_{ij}t^{i}w^{j}\nonumber
	\\
	&=&1+a_{10} t+a_{20} t^2+a_{11} t w+a_{30} t^3+a_{21} t^2 w+a_{12} t w^2+a_{03}w^3+a_{40}t^4+a_{31} t^3 w+a_{22} t^2 w^2+a_{13} t w^3 +\nonumber
	\\& &a_{04} w^4+a_{50} t^5+a_{41} t^4 w+ ..., \label{Pexapnsion}
\end{eqnarray}
where $t=1-\tilde{T}$, $w=\tilde{V}-1$. Note that, for convenience and generality, we have used $a_{ij}$ instead of $R, B, D, H, I, G...$ in (\ref{PressureGeneral}) or (\ref{Pyuehua}) to represent the coefficient of $t^i w^j$ term, while it is easily found that $a_{10}=-R,~a_{11}=-B,~a_{03}=D,~a_{20}=H,~a_{12}=-I,~a_{04}=G$. On the other hand, after using Maxwell's equal area law, we obtain following equations
\begin{eqnarray}
	& &a_{03}( w_l^3-w_s^3)+a_{11}t(  w_l-  w_s)+a_{04}( w_l^4- w_s^4)+a_{12}t(  w_l^2-  w_s^2)+a_{13}t(  w_l^3-  w_s^3)+\nonumber
	\\& &a_{21}t^2(  w_l-  w_s)+a_{22}t^2(  w_l^2- w_s^2)+a_{31}t^3(  w_l- w_s)+a_{41} t^4( w_l- w_s)+...=0,\nonumber
	\\& &\frac{3}{4}a_{03}(w_l^4-w_s^4)+\frac{1}{2}a_{11}t(w_l^2-w_s^2)+\frac{4}{5}a_{04}(w_l^5-w_s^5)+\frac{2}{3}a_{12} t(w_l^3-w_s^3)+\frac{3}{4}a_{13} t(w_l^4-w_s^4)+\nonumber
	\\& &\frac{1}{2}a_{21} t^2(w_l^2-w_s^2)+\frac{2}{3}a_{22} t^2(w_l^3-w_s^3)+\frac{1}{2}a_{31} t^3(w_l^2-w_s^2)+\frac{1}{2}a_{41} t^4(w_l^2-w_s^2)+...=0.\label{equation1}
\end{eqnarray}
Obviously, if $a_{03}$ and $a_{11}$ are not zero, leading terms of the above equations are
\begin{eqnarray}
	& &a_{03}( w_l^3-w_s^3)+a_{11}t(  w_l-  w_s)=0,~~\frac{3}{4}a_{03}(w_l^4-w_s^4)+\frac{1}{2}a_{11}t(w_l^2-w_s^2)=0,\label{equation}
\end{eqnarray}
and the corresponding analytical solution is
\begin{eqnarray}
	w_s=-\sqrt{-\frac{a_{11}}{a_{03}}}t^{1/2},~~ w_l=\sqrt{-\frac{a_{11}}{a_{03}}}t^{1/2},
\end{eqnarray}
which can be considered as the leading term in general solution. Considering the first order correction terms of equation (\ref{equation1}),
	one can get
	\begin{eqnarray}
		& &a_{03}( w_l^3-w_s^3)+a_{11}t(  w_l-  w_s)+a_{04}( w_l^4- w_s^4)+a_{12}t(  w_l^2-  w_s^2)=0,
		\nonumber
		\\& &\frac{3}{4}a_{03}(w_l^4-w_s^4)+\frac{1}{2}a_{11}t(w_l^2-w_s^2)+\frac{4}{5}a_{04}(w_l^5-w_s^5)+\frac{2}{3}a_{12} t(w_l^3-w_s^3)=0,
	\end{eqnarray}
	so one can easily obtain the general solution to the next leading terms as
	\begin{eqnarray}
		w_s=-\sqrt{-\frac{a_{11}}{a_{03}}}t^{1/2}+\left(\frac{6a_{04}a_{11}}{15a^2_{03}}-\frac{a_{12}}{3a_{03}}\right)t, ~~ w_l=\sqrt{-\frac{a_{11}}{a_{03}}}t^{1/2}+\left(\frac{6a_{04}a_{11}}{15a^2_{03}}-\frac{a_{12}}{3a_{03}}\right)t, ~~\label{vAppendix}
	\end{eqnarray}
	and use the same procedure, we can obtain the general solution up to the third leading term as
	\begin{eqnarray}
	w_s&=&-\sqrt{-\frac{a_{11}}{a_{03}}}t^{1/2}+\left(\frac{6a_{04}a_{11}}{15a^2_{03}}-\frac{a_{12}}{3a_{03}}\right)t \nonumber \\
&+&\sqrt{-\frac{a_{11}}{a_{03}}}\frac{100a_{03}a_{04}a_{11}a_{12}-84a^4_{04}a^2_{11}+75a_{03}a^2_{11}+25a^2_{03}(3a_{03}a_{21}-a^2_{12}-3a_{11}a_{13})}{150a^2_{03}a_{11}}t^{3/2}, \nonumber
	 \\w_l&=&\sqrt{-\frac{a_{11}}{a_{03}}}t^{1/2}+\left(\frac{6a_{04}a_{11}}{15a^2_{03}}-\nonumber\frac{a_{12}}{3a_{03}}\right)t \nonumber \\
&-&\sqrt{-\frac{a_{11}}{a_{03}}}\frac{100a_{03}a_{04}a_{11}a_{12}-84a^4_{04}a^2_{11}+75a_{03}a^2_{11}+25a^2_{03}(3a_{03}a_{21}-a^2_{12}-3a_{11}a_{13})}{150a^2_{03}a_{11}}t^{3/2}, \label{wthird}
	\end{eqnarray}
which is a little complicated. However, we will prove that the terms higher than $t$ in the general solution do not affect the precise coefficients of $R_N$ in (\ref{RNfinall}) proportional to $t^{-2}$ and $t^{-3/2}$.
	
For convenience, we write $R_N$ in (\ref{RN}) with $\tilde{P}$, $t$ and $w$
\begin{eqnarray}
	R_N=\frac{1}{2}-\frac{(t-1)^2\left(\frac{\partial^2 \tilde{P}}{\partial t\partial w}\right)^2}{2\left(\frac{\partial \tilde{P}}{\partial w}\right)_t^2}+\frac{(t-1)^2\left(\frac{\partial^3 \tilde{P}}{\partial t^2\partial w}\right)}{\left(\frac{\partial \tilde{P}}{\partial w}\right)_t}\label{app RN}.
\end{eqnarray}
From (\ref{Pexapnsion}), we easily find
\begin{eqnarray}
	 \left(\frac{\partial \tilde{P}}{\partial w}\right)_t&=&\left(\frac{\partial \tilde{P}}{\partial \tilde{V}}\right)_{\tilde{T}}=3 a_{03} w^2+4 a_{04} w^3
	+a_{11} t+2 a_{12} t w+3 a_{13} t w^2+a_{21} t^2+2 a_{22} t^2 w+a_{31} t^3+a_{41} t^4+...,\nonumber
	\\\frac{\partial^2 \tilde{P}}{\partial t\partial w}&=&\frac{\partial^2 \tilde{P}}{\partial \tilde{T}\partial \tilde{V}}=a_{11}+2 a_{12} w+3 a_{13} w^2
	+2 a_{21} t+4 a_{22} t w+3 a_{31} t^2+4 a_{41} t^3+...,\nonumber
	\\\frac{\partial^3 \tilde{P}}{\partial t^2\partial w}&=&\frac{\partial^3 \tilde{P}}{\partial \tilde{T}^2\partial \tilde{V}}=2 a_{21}+4 a_{22} w+6 a_{31} t
	+12 a_{41} t^2+...\label{partial}
\end{eqnarray}
For further calculations, we simplify the $w$ in (\ref{wthird}) as $w=xt^{1/2}+yt+zt^{3/2}+...$, and hence obtain
\begin{eqnarray}
\left(\frac{\partial \tilde{P}}{\partial w}\right)_t^{-1}&=&\frac{1}{\left(3 a_{03} x^2+a_{11}\right)t+2 \left(3 a_{03} x y+2 a_{04} x^3+a_{12} x\right)t^{3/2}+...} \nonumber\\
&=&\frac{1}{ \left(3 a_{03} x^2+a_{11}\right)t}-\frac{2 \left(3 a_{03} x y+2 a_{04} x^3+a_{12} x\right)}{ \left(3 a_{03} x^2+a_{11}\right)^2t^{1/2}}+...,\label{Inverse}
\end{eqnarray}
therefore, for the second term of $R_N$ in (\ref{app RN}), i.e. $R_{N2}\equiv-\left[(t-1)^2\left(\frac{\partial^2 \tilde{P}}{\partial t\partial w}\right)^2\right]/\left[2\left(\frac{\partial \tilde{P}}{\partial w}\right)_t^2\right]$, its expansion around $t=0$ is
	\begin{eqnarray}
		R_{N2}&=&-\frac{1}{2}(t-1)^2\left(a_{11}+2 a_{12} w+3 a_{13} w^2+2 a_{21} t+4 a_{22} t w+3 a_{31} t^2+4 a_{41} t^3+...\right)^2\nonumber
		\\& &\times\left[\frac{1}{ \left(3 a_{03} x^2+a_{11}\right)^2t^2}-\frac{4 \left(3 a_{03} x y+2 a_{04} x^3+a_{12} x\right)}{ \left(3 a_{03} x^2+a_{11}\right)^3t^{3/2}}+...\right]\nonumber\\
&=&-\frac{a_{11}^2}{2(3a_{03}x^2+a_{11})^2}\frac{1}{t^{2}}+\left[\frac{2a_{11}^2(3a_{03}xy+2a_{04}x^3+a_{12}x)}{(3a_{03}x^2+a_{11})^3}-\frac{2a_{11}a_{12}x}{(3a_{03}x^2+a_{11})^2}\right]\frac{1}{t^{3/2}}+O(t^{-1}),\label{RN2}
	\end{eqnarray}
 while the third term of $R_N$ in (\ref{app RN}), i.e. $R_{N3}\equiv\frac{(t-1)^2\left(\frac{\partial^3 \tilde{P}}{\partial t^2\partial w}\right)}{\left(\frac{\partial \tilde{P}}{\partial w}\right)_t}$ is expanded around $t=0$ as
	\begin{eqnarray}
	R_{N3}&=&(t-1)^2(2 a_{21}+4 a_{22} w+6 a_{31} t
		+12 a_{41} t^2+...) \times\left[\frac{1}{ \left(3 a_{03} x^2+a_{11}\right)t}-\frac{2 \left(3 a_{03} x y+2 a_{04} x^3+a_{12} x\right)}{ \left(3 a_{03} x^2+a_{11}\right)^2t^{1/2}}+...\right]\nonumber
		\\&=&\frac{2 a_{21}}{ \left(3 a_{03} x^2+a_{11}\right)}t^{-1}+O(t^{-1/2})=-(a_{21}/a_{11}) t^{-1}+O(t^{-1/2}),\label{RN3}
	\end{eqnarray}
where $x=\pm \sqrt{-\frac{a_{11}}{a_{03}}}$ from (\ref{wthird}) have been used in the final step. From above results in $R_{N2}$ and $R_{N3}$, we obviously find that higher terms like $t^{3/2}$ in $w$ do not affect the precise coefficients of $R_N$ proportional to $t^{-2}$ and $t^{-3/2}$ in (\ref{RNfinall}). Therefore, after substituting (\ref{RN2}) and (\ref{RN3}) into (\ref{app RN}), and using $y=\frac{6a_{04}a_{11}}{15a^2_{03}}-\frac{a_{12}}{3a_{03}}$ from (\ref{wthird}), we finally obtain the divergence behaviour of $R_N$ around $t=0$
\begin{eqnarray}
R_N\left(SBH\right)=-\frac{1}{8 t^2}+\sqrt{-\frac{a_{11}}{a_{03}}} \frac{(5 a_{03} a_{12}-2 a_{04} a_{11})}{10  (a_{03} a_{11})t^{3/2}}+O(t^{-1}),\nonumber
	\\R_N\left(LBH\right)=-\frac{1}{8 t^2}-\sqrt{-\frac{a_{11}}{a_{03}}} \frac{(5 a_{03} a_{12}-2 a_{04} a_{11})}{10  (a_{03} a_{11})t^{3/2}}+O(t^{-1}),\label{RNgeneric}
\end{eqnarray}
which is valid for generic black holes with nonzero parameters $a_{03}$ and $a_{11}$. After using $a_{10}=-R,~a_{11}=-B,~a_{03}=D,~a_{20}=H,~a_{12}=-I,~a_{04}=G$, we can easily find that the results in (\ref{RNgeneric}) are consistent with those in  (\ref{RNfinall}). Furthermore, from these results, we strictly prove that the terms higher than $t$ of $w$ or $\tilde{V}$ do not affect the coefficients of $t^{-2}$ and $t^{-3/2}$ terms in $R_N$, while the $R_{N3}$ term also does not affect since the leading term in $R_{N3}$ is proportional to $1/t$. Therefore, the results in (\ref{RNfinall}) is solid in the case of generic black holes with nonzero $B$ and $D$.

\section{Checking the universality in dRGT massive gravity}\label{4}
 Note that it is worth mentioning that the dRGT massive gravity is a well-defined modified gravity with massive graviton. Moreover, the $P$-$V$ criticality in extended phase space of black holes in dRGT massive gravity has been shown to have special behavior~\cite{Xu:2015rfa}, i.e. $P$-$V$ criticality appears not only in the spherical horizon topology with $k=1$ but also in the Ricci flat case with $k=0$ and hyperbolic case with $k=-1$. Particularly, as far as we know, this special behavior only exists in the dRGT massive gravity among all gravitational theories. Therefore, it is interesting to investigate divergence behavior of thermodynamic curvature scalar $R_N$ and check the universality in the dRGT massive gravity, particularly with Ricci flat or hyperbolic horizon topology case. In this section, we mainly use the method mentioned in Sec~\ref{four} to investigate thermodynamic curvature scalar $R_N$ in the four-dimensional dRGT massive gravity case to check the universality.

In four-dimensional spacetime, the pressure $P$ in the extended phase space of black hole in dRGT massive gravity is \cite{Xu:2015rfa}
\begin{eqnarray}	 P=\left(\dfrac{T}{2}-\dfrac{c_0c_1m^2}{8\pi}\right)\dfrac{1}{r_h}-\left(\dfrac{k}{8\pi}+\dfrac{c^2_0c_2m^2}{8\pi}\right)\dfrac{1}{r^2_h}+\dfrac{8\pi Q^2}{V^2_2}\dfrac{1}{r^4_h},\label{3.1(4d)}
\end{eqnarray}
where $r_h$ is the horizon radius of black hole, $c_i(i=0,1,2)$ are constants, $m$ is massive graviton parameter, $k$ characterizes the horizon curvature, $V_2$ is the area of space spanned by coordinates with horizon topology $k=0, \pm 1$, $Q$ is related to the charge of black hole.

For convenience, the pressure in (\ref{3.1(4d)}) is rewritten as
\begin{eqnarray}
P=\dfrac{w_1}{r_h}+\dfrac{w_2}{r^2_h}+\dfrac{w_4}{r^4_h}, \label{4dPressure}
\end{eqnarray}
where
\begin{eqnarray}
w_1&=&\dfrac{T}{2}-\dfrac{c_0c_1m^2}{8\pi},~~w_2=-\left(\dfrac{k}{8\pi}+\dfrac{c^2_0c_2m^2}{8\pi}\right),~~w_4=\dfrac{8\pi Q^2}{V^2_2}. \label{NewParameters}
\end{eqnarray}
After using $\left[\left(\dfrac{\partial P}{\partial V}\right)_T\right]_c=\left[\left(\dfrac{\partial ^2P}{\partial V^2}\right)_T\right]_c=0$, the critical point has been obtained
\begin{eqnarray}
r_{hc}&=&\sqrt{-\dfrac{6w_4}{w_2}}~~, w_{1c}=-\dfrac{4}{3}w_2\sqrt{-\dfrac{w_2}{6w_4}},~~P_c=\dfrac{w^2_2}{12w_4}.
\end{eqnarray}
Note that, $R_N$ is a dimensionless parameter in (\ref{RN}). Therefore, for convenience of calculations in the following, we rewrite pressure (\ref{4dPressure}) in the reduced parameter space as
\begin{eqnarray}
\tilde{P}=\dfrac{8(\tilde{T}-2\tilde{F})}{3\tilde{r}_h\left(1-2\tilde{F}\right)}-\dfrac{2}{\tilde{r}_h^2}+\dfrac{1}{3\tilde{r}_h^4}, \label{4dPreduced}
\end{eqnarray}
where $\tilde{r}_h=\dfrac{r_h}{r_{hc}}$, and $\tilde{F}=\dfrac{F}{T_c}=\dfrac{c_0c_1m^2}{8\pi T_c}$ with $F=\dfrac{c_0c_1m^2}{8\pi}$. On the other hand, $R_N$ in (\ref{RN}) is calculated through the pressure function $P(T,V)$, where $V=\dfrac{V_2}{3}r^3_h$ is the thermodynamic volume of black hole. Hence we further rewrite the reduced pressure (\ref{4dPreduced}) with reduced volume $\tilde{V}=\dfrac{V}{V_c}$ as
\begin{eqnarray}		
\tilde{P}=\dfrac{8}{3}\dfrac{\tilde{T}-2\tilde{F}}{\tilde{V}^{1/3}\left(1-2\tilde{F}\right)}-\dfrac{2}{\tilde{V}^{2/3}}+\dfrac{1}{3\tilde{V}^{4/3}},\label{4dPreduced2}
\end{eqnarray}
 In addition, entropy of black hole has been found to be $S=\dfrac{V_2}{4}r^2_h$ in Ref~\cite{Xu:2015rfa}, and hence heat capacity $C_V$ of black hole is easily found to be zero if one fixes the thermodynamic volume $V=\dfrac{V_2}{3}r^3_h$ of black hole. The expansion behavior of $\tilde{P}$ around the critical point is easily obtained
\begin{eqnarray}
\tilde{P}=1+\dfrac{8T_c}{6F-3T_c}t-\dfrac{8T_c}{9\left(2F-T_c\right)}wt-\dfrac{4}{81}w^3-\dfrac{16T_c}{27\left(T_c-2F\right)}w^2t+\dfrac{25}{243}w^4+...,
\end{eqnarray}
and the coefficients of (\ref{Pyuehua}) in this case are directly read as $B=\dfrac{8T_c}{9\left(2F-T_c\right)}, D=-\dfrac{4}{81}, G=\dfrac{25}{243},I=\dfrac{16T_c}{27\left(T_c -2F\right)}$.

	
After substituting those values of $B,~D,~I$ and $G$ into thermodynamic curvature scalar $R_N$ in (\ref{RNfinall}), we can obtain the divergence behavior of $R_N$ around the critical point when $t$ approaches to $0$
\begin{eqnarray}
R_N\left(SBH\right)=-\frac{1}{8 t^2}+\frac{\sqrt{T_c/\left(T_c-2 F\right)}}{2\sqrt{2} t^{3/2}}+O(t^{-1}),
~~R_N\left(LBH\right)=-\frac{1}{8 t^2}-\frac{\sqrt{T_c/\left(T_c-2 F\right)}}{2\sqrt{2} t^{3/2}}+O(t^{-1}),
\label{RNseries2-1}
\end{eqnarray}
for small (SBH) and large (LBH) black hole cases, respectively. Obviously, this divergence behavior is still characterized by a dimensionless constant $-1/8$ from the coefficient in the first divergent term, and the dimensionless constant is indeed independent of graviton mass and horizon topology of black holes. It should be pointed that, we can use other analytical method like Ref~\cite{Spallucci:2013osa} to obtain the divergence behavior of $R_N$ in this four dimensional dRGT massive gravity case, which is also calculated in details and found to be consistent with (\ref{RNseries2-1}) in appendix~\ref{AppendixA}.

\section{Conclusions and discussions}

In this paper, motivated from the divergence behavior of Ruppeiner thermodynamic curvature scalar $R_N$ at critical point in the extended phase space of a charged AdS black hole, we further propose a method to investigate $R_N$ around the critical point in generic black holes. We demonstrate that the dimensionless constant $-1/8$ characterized the divergence behavior of $R_N$ at critical point indeed expresses a universal property in generic black holes, i.e. black holes exist $P-V$ criticality and have nonzero $B$ and $D$. Furthermore, we also use our method to check this universality in the four-dimensional dRGT massive gravity case, since dRGT massive gravity has manifested some special properties different from other $P-V$ criticality in the extended phase space of asymptotic AdS black holes. Our results show that this universality is still satisfied in the four-dimensional dRGT massive gravity case, where the dimensionless constant is still $-1/8$ and independent of graviton mass and horizon topology of black holes.

It is impressive that the thermodynamic curvature scalar $R_N$ expresses a universal divergence behavior at the critical point. However, until now we still have little knowledge about the underlying physical origin and the consequences of this universal divergence behavior. Hence, it is interesting to further investigate this divergence behavior. Note that, during demonstrating this universal divergence behavior in generic black holes, we have made some assumptions, i.e. $C_V$ is zero or constant. If we relax these assumptions, one may obtain new results beyond the mean field theory, in which the dimensionless constant may not be $-1/8$. This picture is somehow consistent with the phenomenon of the normal-superconductor phase transition in liquid Helium, where there is a divergent behavior for the heat capacity $C_V$ when temperature approaches the critical temperature of liquid Helium. Therefore, finding such black hole spacetime is also an open issue. The AdS/CFT correspondence may give us some insights into this issue, since one has found that charged asymptotic AdS black holes with complex scalar field undergo a normal-superconductor phase transition on their conformal boundaries. Therefore, further study of the divergence behavior of thermodynamic curvature scalar $R_N$ in modified gravities with complex scalar field is a potential clue, for example investigations in Einstein-Horndeski gravity. On the other hand, if $a_{11}$ or $a_{03}$ is zero, which may exist in some black hole cases like \cite{Dolan:2013ft}, the solutions of $\omega_l$ and $\omega_s$ in (\ref{vAppendix}) will be changed from our demonstration. We easily find that the critical exponent $\beta$ will be affected as well, and hence the corresponding results are underlying beyond the mean field theory, which are interesting and under investigation in the future work. Finally, from the $P$-$V$ criticality in the extended phase space of asymptotic AdS black holes, the large and small phases of black holes can be considered as the corresponding gas and liquid phases of van der Waals system. As we know, the van der Waals system has a third phase-solid phase, and hence whether the black hole system also has $solid$ phase is an interesting topic, which may also illuminate our new understanding of black hole microstructure.

\section{Acknowledgements}

Y.P. Hu thanks a lot for the discussions with Profs. Shao-Wen Wei, Hai-Qing Zhang and Dr. Yihao Yin. This work is supported by National Natural Science Foundation of China (NSFC) under grant Nos. 12175105, 11575083, 11565017, and Top-notch Academic Programs Project of Jiangsu Higher Education Institutions (TAPP).

\appendix

\section{Other analytical method investigating $R_N$ in the four-dimensional dRGT massive gravity case} \label{AppendixA}
In this appendix, we mainly use the analytical method proposed in Ref \cite{Spallucci:2013osa} to investigate $R_N$ in the four-dimensional dRGT massive gravity case. Here, a key point is that one can obtain the analytical relationship between the temperature and the volume at the transition point, i.e.$V_l(T)$ or $V_s(T)$. Note that, it is a little difficult to directly obtain this analytical relationship, one often first obtains the entropy $S(T)$ of these two black holes phases by following the method in \cite{Spallucci:2013osa}.

Similar to the Maxwell's equal area law $\tilde{P}^*\left(\tilde{V}_s-\tilde{V}_l\right)=\int^s_l\tilde{P}d\tilde{V}$ in the $P$-$V$ phase diagram, where $\tilde{P}^*$ is the pressure of coexistence curves, i.e. $\tilde{P}^*=\tilde{P}_s=\tilde{P}_l$, one can also obtains the Maxwell's equal area law in the reduced $\tilde{T}$-$\tilde{S}$ phase diagram as \cite{Spallucci:2013osa}
\begin{eqnarray}
	\tilde{T}^*\left(\tilde{S_s}-\tilde{S_l}\right)&=&\int^s_l\tilde{T}d\tilde{S},
\end{eqnarray}
where $\tilde{T}^*=\tilde{T}_s=\tilde{T}_l$ and $\tilde{S}=S/S_c$. In the dRGT massive gravity, $\tilde{w}_1$ can be considered as an effective temperature from (\ref{NewParameters}), and it is more convenient to use $\tilde{w}_1$ instead of $\tilde{T}$ in the discussions~\cite{Xu:2015rfa}. Therefore, for the above Maxwell's equal area law in the $\tilde{T}$-$\tilde{S}$  phase diagram, we further easily obtain
\begin{eqnarray}
	\tilde{w}_1^*\left(\tilde{S}_s-\tilde{S}_l\right)&=&\int^s_l\tilde{w}_1d\tilde{S}, \label{NewMaxwell}
\end{eqnarray}
where $\tilde{w}_1^*=\tilde{w}_{1s}=\tilde{w}_{1l}$, and the reduced parameter $\tilde{w}_1$ is defined as $\tilde{w}_1=\dfrac{\tilde{T}-2F/T_c}{1-2F/T_c}=\dfrac{3}{4\sqrt{\tilde{S}}}\left(1+\dfrac{1}{2} \tilde{P} \tilde{S} -\dfrac{1}{6\tilde{S}}\right)$.  From this equation (\ref{NewMaxwell}), we get
\begin{eqnarray}
\tilde{w}_1^*=\dfrac{3/2-1/4\sqrt{\tilde{S}_s \tilde{S}_l}+\tilde{P} \left(\tilde{S}_s +\tilde{S}_l +\sqrt{\tilde{S}_s \tilde{S}_l}\right)/4}{\sqrt{\tilde{S}_s}+\sqrt{\tilde{S}_l}},~~
\tilde{w}_1\left(\tilde{S}_s\right)=\tilde{w}_1\left(\tilde{S}_l\right).
\end{eqnarray}
Then the entropy of small and large phases of black hole are solved as
\begin{eqnarray}
\tilde{S}_s&=&\frac{\left(\sqrt{3-\sqrt{\tilde{P}}}-\sqrt{3-3\sqrt{\tilde{P}}}\right)^2}{2\tilde{P} },~~ \tilde{S}_l=\frac{\left(\sqrt{3-\sqrt{\tilde{P}}}+\sqrt{3-3\sqrt{\tilde{P}}}\right)^2}{2\tilde{P} }, \label{4DEntropyT}
\end{eqnarray}
where $\tilde{P}=\left[1-2 \cos \left(\frac{1}{3} \cos ^{-1}\left(1-\frac{\left(\tilde{T}-2 F/T_c\right)^2}{\left(1-2 F/T_c\right)^2}\right)+\frac{1}{3}\pi \right)\right]^2$.
By using the relationship between reduced entropy and reduced thermodynamic volume of black hole $\tilde{S}=\tilde{V}^{2/3}$, we finally obtain relationship between $\tilde{V}$ and $\tilde{T}$
\begin{eqnarray}
	\tilde{V}_s&=&\frac{\left(\sqrt{3-\sqrt{\tilde{P}}}-\sqrt{3-3\sqrt{\tilde{P}}}\right)^3}{2 \sqrt{2}\tilde{P} \sqrt{\tilde{P}} },~~ \tilde{V}_l=\frac{\left(\sqrt{3-\sqrt{\tilde{P}}}+\sqrt{3-3\sqrt{\tilde{P}}}\right)^3}{2 \sqrt{2}\tilde{P} \sqrt{\tilde{P}} }. \label{VTrelation}
\end{eqnarray}

On the other hand, after substituting (\ref{4dPreduced2}) into (\ref{RN}), we easily obtain the explicit expression of thermodynamic curvature scalar $R_N$ of black hole spacetime in the four-dimensional dRGT massive gravity case
\begin{eqnarray}
	R_N=\frac{\left[F \left(-6 \tilde{V}^{2/3}+4 \tilde{V}+2\right)+T_c \left(3 \tilde{V}^{2/3}-1\right)\right] \left[F \left(-6 \tilde{V}^{2/3}+4 \tilde{V}+2\right)+T_c \left(-4 \tilde{T} \tilde{V}+3 \tilde{V}^{2/3}-1\right)\right]}{2 \left[F \left(6 \tilde{V}^{2/3}-4 \tilde{V}-2\right)+T_c \left(2 \tilde{T} \tilde{V}-3 \tilde{V}^{2/3}+1\right)\right]^2}.\label{4dFinalRN}
\end{eqnarray}
after substituting (\ref{VTrelation}) into thermodynamic curvature scalar $R_N$ in (\ref{4dFinalRN}), we can finally obtain the divergence behavior of $R_N$ around the critical point
\begin{eqnarray}
	R_N\left(SBH\right)=-\frac{1}{8 t^2}+\frac{\sqrt{T_c/\left(T_c-2 F\right)}}{2\sqrt{2} t^{3/2}}+O(t^{-1}),~~R_N\left(LBH\right)=-\frac{1}{8 t^2}-\frac{\sqrt{T_c/\left(T_c-2 F\right)}}{2\sqrt{2} t^{3/2}}+O(t^{-1}).
	\label{RNseries2}
\end{eqnarray}
Obviously, the $R_N$ obtained in (\ref{RNseries2}) is consistent with that of our method in (\ref{RNseries2-1}).

\end{document}